\documentclass[aps,pre,twocolumn,unsortedaddress]{revtex4}
\usepackage[utf8]{inputenc}
\usepackage{graphicx}

\begin{document}

\title{Ordered domains in sheared dense suspensions: the link to viscosity and the disruptive effect of friction}
\author{Abhay Goyal}
\affiliation{Infrastructure Materials Group, Engineering Laboratory, National Institute of Standards and Technology}
\affiliation{Dept of Physics, Institute of Soft Matter Synthesis and Metrology, Georgetown University}

\author{Emanuela Del Gado}
\affiliation{Dept of Physics, Institute of Soft Matter Synthesis and Metrology, Georgetown University}
\author{Scott Z. Jones}
\affiliation{Infrastructure Materials Group, Engineering Laboratory, National Institute of Standards and Technology}
\author{Nicos S. Martys}
\affiliation{Infrastructure Materials Group, Engineering Laboratory, National Institute of Standards and Technology}

\begin{abstract}
    
    Monodisperse suspensions of Brownian colloidal spheres crystallize at high densities, and ordering under shear has been observed at densities below the crystallization threshold. We perform large-scale simulations of a model suspension containing over $10^5$ particles to quantitatively study the ordering under shear and to investigate its link to the rheological properties of the suspension. We find that at high rates, for $Pe>1$, the shear flow induces an ordering transition that significantly decreases the measured viscosity. This ordering is analyzed in terms of the development of layering and planar order, and we determine that particles are packed into hexagonal crystal layers (with numerous defects) that slide past each other. By computing local $\psi_6$ and $\psi_4$ order parameters, we determine that the defects correspond to chains of particles in a square-like lattice. We compute the individual particle contributions to the stress tensor and discover that the largest contributors to the shear stress are primarily located in these lower density, defect regions. The defect structure enables the formation of compressed chains of particles to resist the shear, but these chains are transient and short-lived. The inclusion of a contact friction force allows the stress-bearing structures to grow into a system-spanning network, thereby disrupting the order and drastically increasing the suspension viscosity.
\end{abstract}

\maketitle

\section{Introduction}
Suspensions are ubiquitous in nature and play an important role in a wide variety of environmental and technical processes.  Examples include concrete, pharmaceuticals, blood, and more. Dilute suspensions typically exhibit liquid-like behavior, but when the volume fraction of suspended particles is high, complex rheological properties arise \cite{Royer2016,Guy2019,Ness2018,Pradeep2021}. Understanding and controlling this behavior is of academic interest and crucial for many industrial applications.  

Colloidal suspensions are known to order under shear, giving rise to shear thinning properties \cite{Foss2000,Ackerson1990,Wu2009,Stevens1993,Lee2018,Rastogi1996,Kulkarni2009,Mabhoff2020}. Dense hard sphere suspensions will also order at rest due to thermal motion, given enough time, forming fcc or hcp lattices \cite{Pusey1986,Cheng2001,Volkov2002}. The dynamics of this crystallization process are complex, depending strongly on the volume fraction of particles, particle size distribution, boundary conditions, and the applied shear rate \cite{Volkov2002,Lander2013,Richard2015,Mabhoff2020,Holmqvist2005}.

In addition, at high densities many suspensions exhibit shear thickening properties \cite{Rathee2020,Royer2016,Cwalina2014,Pradeep2021,Morris2020}. This is generally associated with cluster formation and contact forces between particles. At higher shear rates, more frequent contact between particles gives rise to large microscopic forces and a high macroscopic viscosity. This rise in viscosity is called continuous shear thickening (CST) or discontinuous shear thickening (DST), depending on how rapidly the viscosity rises with shear rate. There is substantial evidence that frictional forces, due to particle roughness, play a key role in the shear thickening phenomenon, with DST being attributed to a sharp rise in the number of frictional contacts \cite{Wyart2014,Singh2020,Morris2018,Royer2016,Pradeep2021,Guy2019}.

Using a model monodisperse suspension of hard spheres, we investigated the ordering transition for a system in the liquid-crystal coexistance regime of the equilibrium phase diagram. We analyzed the types of ordered states that arise under shear, as well as how the ordered domains nucleate and grow. By computing the distribution of stresses within the suspension, we discovered that the largest contributions to the shear stress come from the defects in the ordered state. The defect structure enables the formation of compressed chains resisting the flow. While these chains are short-lived in the frictionless system, the addition of contact friction between particles disrupts the order and allows the stress-carrying structures to grow into a system-spanning network. 

This paper is organized as follows. First, we give a detailed introduction to the model and describe the simulation and analysis performed in section \ref{sec:methods}. Next, we discuss the ordering transition the sheared suspension undergoes and its effect on viscosity in section \ref{sec:ordering}. In section \ref{sec:orderingAnalysis}, we quantitatively analyze the ordering and discuss how it develops over time, as the suspension is sheared. The formation of defects and analysis of their structure is contained in section \ref{sec:defects}. The microstructure analysis is then related to the mechanics by analyzing the local particle stresses in section \ref{sec:stresses}. Finally, we comment on the role of friction in the disruption of ordering and redistribution of stresses in section \ref{sec:friction}.

\section{Methods}
\label{sec:methods}

\subsection{Simulation details}
We performed discrete element method (DEM) simulations of a suspension of monodisperse, spherical particles. The suspension contained 108,000 particles in a periodic simulation box with the particle volume fraction ranging from $\phi = 54\,\% - 58\,\%$, with most of the analysis presented being focused on the $\phi=54\,\%$ case. This is a range of $\phi$ where we observe substantial ordering and which spans CST and DST regimes as observed in simulations with friction \cite{Mari2014,Seto2013}. While below the static jamming volume fraction, the suspensions can shear jam at high shear rates when $\phi \gtrsim 56\,\%$ \cite{Mari2014,Seto2013}. The solvent was treated implicitly, with a viscosity of $\eta_0=1$ in reduced units of $\sqrt{m\varepsilon}/d$. The particle mass $m$ and diameter $d$, as well as the simulation energy scale $\varepsilon$, were also set to 1. Note that the particles have finite mass because we are solving Newton's equations of motion, which include inertia. However, we have ensured we are in the overdamped limit by running simulations with suspension viscosity increased up to $\eta_0=10$ and finding that its value does not affect our results. 

All simulations were performed with LAMMPS (Large-scale Atomic/Molecular Massively Parallel Simulator) \cite{Plimpton1995}. While generally known for molecular dynamics simulations, the LAMMPS code has specialized modules that allow for the modeling of hard sphere suspensions of Brownian particles. For this application, LAMMPS utilizes a simple physics-based DEM model that ignores the detailed flow behavior of the suspension solvent in exchange for computational efficiency. Hydrodynamic interactions between spheres are largely controlled by lubrication forces, as discussed in the next section. Shear flow was imposed along the $x$ direction (with gradient along $z$) based on the Lees-Edwards boundary condition, with an additional drag force similar to a Couette device. The robustness of this approach improves with increasing volume fraction of spheres as a less detailed knowledge of flow of the background fluid is needed. Indeed, it has been found that at volume fractions of approximately $40\,\%$ and higher, the flows produced are reasonably consistent with fully detailed simulations \cite{Feys2022}. This is due to the fact that, at higher volume fractions, the surfaces of the solid inclusions are close enough such that the lubrication forces dominate over the long-range hydrodynamics of the background fluid \cite{Ball1997}.

In addition to the lubrication forces, a short-ranged Yukawa potential was used to give the particles an exponentially decaying steric repulsion following $U_{\rm Yukawa}=\frac{A}{\kappa}e^{-\kappa h}$. Here $h$ is the separation between particle surfaces, while $A=1000\,\varepsilon/d$ and $\kappa=100\,d^{-1}$ are set so that the energy is $10\,\varepsilon$ at contact and falls below $k_BT$ at a surface separation of $h=0.02\,d$.  Thermal fluctuations were introduced using a random Brownian force to maintain a finite temperature such that $k_BT=\varepsilon$. This sets the time scale in the simulations as $\tau=\sqrt{\frac{md^2}{k_BT}}$. We tested the model parameters for a slightly more dilute suspension ($\phi=45\,\%$) and found results in excellent agreement with the earlier work of Foss and Brady where they performed simulations of hard spheres with hydrodynamic interactions \cite{Foss2000}.

All shear simulations were run from a disordered initial state. This state was prepared by first placing the particles on a cubic lattice then running a high temperature DEM simulation without any shear. Once the crystal had melted, the system was cooled to $k_BT=1\,\varepsilon$, and the pair correlation was checked to ensure the configuration was amorphous. For all rates considered, a copy of this initial state was sheared until the system reached steady state, as evidenced by steady viscosity, pressure, and energy. To ensure that we remained in the viscous regime rather than inertial (see \cite{Madraki2020} for more on this transition), we performed additional simulations with $\eta_0=10$ and found that the suspension viscosity was unchanged.

\subsection{Lubrication forces}
Hydrodynamic lubrication forces arise due to the relative motion of particles through the solvent. As the particles move, they drag the solvent with them, which then exerts a force on nearby particles. The interactions due to the relative motion of the spheres have been described by Ball and Melrose \cite{Ball1997}, who computed forces ($\vec{F}$) and torques ($\vec{\tau}$) acting on spherical particles as

\begin{figure}
    \centering
    \includegraphics[width=.45\textwidth]{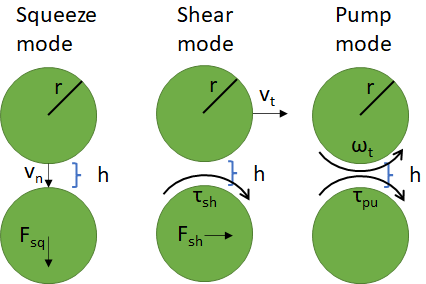}
    \includegraphics[width=.45\textwidth]{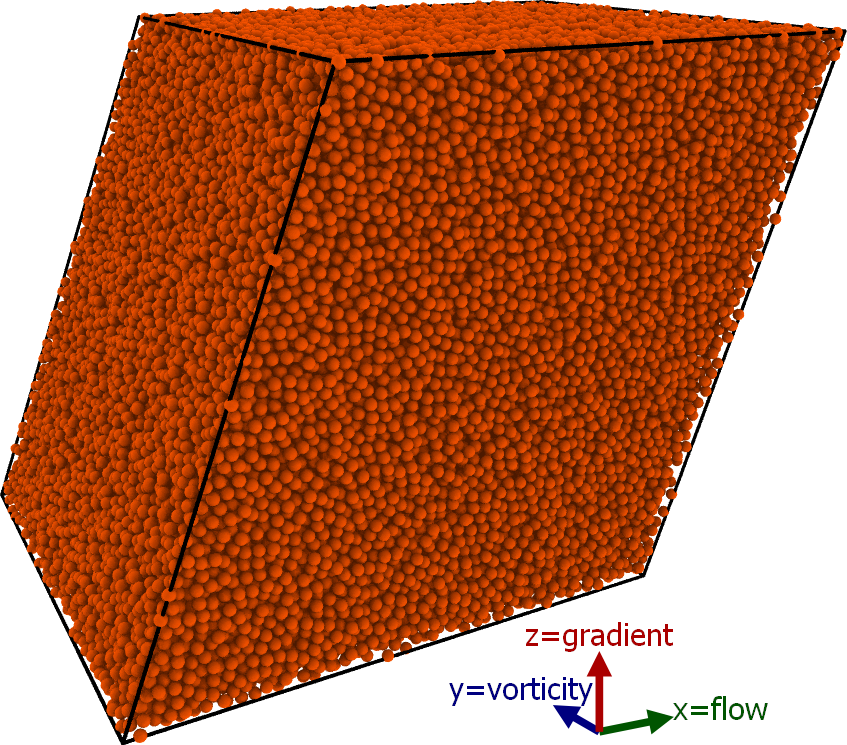}
    \caption{\textbf{Top:} Schematic representation of the different modes of relative motion between spheres in a background solvent \cite{Ball1997}. Each of these modes has a contribution to the lubrication forces and/or torques between the two spheres. The motion of the top sphere drags the solvent and exerts a force/torque on the bottom sphere. \textbf{Bottom:} Snapshot of the whole simulation box containing 108,000 spheres, all interacting via the hydrodynamic lubrication force. The shear is imposed along the $x$ direction, with the velocity gradient along the $z$ direction. }
    \label{fig:lub}
\end{figure}

\begin{equation}
	\vec{F_{sq}} = a_{sq} \vec{v_n} 
\end{equation}
\begin{equation}
	\vec{F_{sh}} = a_{sh} \vec{v_t} 
\end{equation}
\begin{equation}
\vec{\tau_{sh}} =  \vec{r} \times \vec{F_{sh}} 
\end{equation}
\begin{equation}
	\vec{\tau_{pu}} = a_{pu} \vec{\omega_t}
\end{equation}
where the forces and torques denoted by $sq$, $sh$, and $pu$ are due to squeezing, shearing, and pumping, respectively, of the fluid between the spheres (see Fig.~\ref{fig:lub}a for a visual depiction of these modes). The velocity of the spheres relative to each other is split into normal and tangential components as $\vec{v_n}$ and $\vec{v_t}$, where the normal is defined as the direction along the center-center line between the spheres. $\vec{\omega_t}$ is the angular velocity difference in directions tangent to the normal direction. The coefficients for Eq.~1-4, as computed by Ball and Melrose \cite{Ball1997}, are 
\begin{equation}
	a_{sq} = 3\pi \eta_0 r \left[\frac{1}{2h} + \frac{9}{20}\log(\frac{1}{h}) \right] 
\end{equation}
\begin{equation}
	a_{sh} = \pi \eta_0 r \left[\log(\frac{1}{h}) \right] 
\end{equation}
\begin{equation}
	a_{pu} = \pi \eta_0 r^3 \left[\frac{3}{20}\log(\frac{1}{h}) \right]
\end{equation}
where $r$ is the radius of the spheres, $h$ is the surface-surface separation between spheres, and $\eta_0$ is the viscosity of the solvent. 

As one can see from the coefficients, these forces are singular at contact ($h=0$), and this singularity is typically handled in simulations by setting an inner cutoff $h_{in}$. This sets a finite minimum separation such that the lubrication forces use $h_{in}$ in lieu of $h$ when $h<h_{in}$. We tested the dependence of the stress response on this threshold and found that for values of $h_{in}>10^{-4}\,d$ the hydrodynamic stresses were underestimated at high rates. The exact threshold is, of course, dependent on the presence of other particle interactions, and a different cutoff can be appropriate if there are other forces, such as electrostatic repulsion, between particles. For all data presented we used $h_{in}=10^{-4}\,d$.

\subsection{Microstructural analysis}

Several analytic tools were used to quantitatively study the ordering of particles in the suspension. We computed the density profiles $\rho(z)$ and found evidence of particle layering. Steinhardt's bond orientational order parameters were used to characterize the 3D ordering \cite{Steinhardt1983}, but no 3D crystal structure was found (see supporting information). Due to the nature of the order, we focus on 2D correlation functions and order parameters that are computed in layers of particles. 

Generally, all of the results that are presented as functions of $z$, such as the density profiles $\rho(z)$ in Fig.~\ref{fig:densityProfile}, are computed by averaging the plotted property over all particles in a slice of the system along the gradient direction. With the density profile as reference, we use a slice width of $2d$ when averaging the correlation functions and order parameters (i.e. the average is over all particles $i$ such that $\Delta z=|z_i-z| \leq d$). Particles in the same layer are identified by a threshold separation of $\Delta z \leq 0.4d$, which corresponds to the width of the density peaks in Fig.~\ref{fig:densityProfile}. To clarify this, let us consider the case of the correlation function computed as:

\begin{equation}
	g(r_{xy},z) = \frac{L_x L_y}{2\pi r r_{\rm bin} N_{z}^2 } \sum_{i} \sum_{j\neq i} \mathcal{H} \left(\frac{r_{\rm bin}}{2} - |r_{xy}-r_{xy}^{ij}|\right)    
	\label{eq:gr}
\end{equation}
Here the Heaviside step function $\mathcal{H}$ is used to count particles $j$ within a certain distance $r_{xy}$ of particle $i$, with discrete bin size $r_{\rm bin} = 0.05$. The quantity is averaged over all $N_z$ particles within the slice of width $2d$ for better statistics, and the calculation is confined to $ij$ pairs within the same layer.

A similar procedure is followed with the order parameters $\psi_6$ and $\psi_4$. The individual particle values are computed, following \cite{Qi2006}, as:

\begin{equation}
\psi_6^a = \left| \frac{1}{N_n} \sum_{b=1}^{N_n} e^{6i\theta_{ab}} \right|^2
\label{eq:phi6}
\end{equation}
\begin{equation}
\psi_4^a = \left| \frac{1}{N_n} \sum_{b=1}^{N_n} e^{4i\theta_{ab}} \right|^2
\label{eq:phi4}
\end{equation}
The sum is limited to the $N_n$ nearest neighbors of particle $a$ in the same layer, and $\theta_{ab}$ measures the angle of the bond between $a$ and $b$. To obtain $\psi_6(z)$, the individual particle $\psi_6^a$ values are averaged for all particles in a slice of width $2d$ centered at $z$. 

In addition to analysis of the microstructure, we calculate the stress in the suspension to connect the ordering to the mechanics. The stress is calculated from the particle contributions to the stress tensor, following \cite{Thompson2009}, as  

\begin{equation}
    \sigma_{\alpha \beta} = -\frac{1}{V}\sum_{i} \left[ \sum_{j \neq i} \left( \frac{1}{2} F_\alpha^{ij}r_\beta^{ij} \right) - mv_\alpha^i v_\beta^i \right]
    \label{eq:stressTotal}
\end{equation}
where $i$ and $j$ denote particles, $\alpha$ and $\beta$ denote dimensions, $V$ is the system volume, $F^{ij}$ is the force between two particles, $r^{ij}$ is the distance between two particles, and $v$ is the particle velocity. The suspension viscosity is computed from the shear stress $\sigma_{xz}$ and shear rate $\dot{\gamma}$ as $\eta = \sigma_{xz}/\dot{\gamma}$. In Sec.~\ref{sec:stresses}, we discuss how the total stress can be divided into particle stresses to give local information.

\subsection{Friction}
When including frictional forces, we follow the contact model of Mari et al \cite{Mari2014,Seto2013}. The steric repulsion between particles is modeled as a Hookean force, with a normal force of $F_N=kh$. Unlike the Yukawa-like potential used previously, this allows for particle contact (which is the criterion for activating frictional forces). However, to maintain a hard contact, we use a high spring constant of $k=10^4\,\dot{\gamma}\,\sqrt{mk_BT}$. With these parameters, we observe particle overlap $<6\,\%$, comparable to the criteria used in other recent simulations \cite{Morris2018}. The dependence on shear rate arises from the fact that a higher spring constant is needed to limit overlap at higher shear rates, due to collisions happening more frequently and at greater velocities. Alternatively, one could use a fixed, high value of $k$, corresponding to what is required for the highest shear rate, at all rates. However, this requires a smaller timestep to adequately resolve the collisions, and the variable spring constant approach was found to reduce computational cost while producing results that matched the results from the constant $k$ approach \cite{Morris2018}.

The frictional forces act tangentially to the particle contact and also follow a simple Hookean model. Any tangential displacement of particles after making contact, $\Delta r_t$, is acted upon by a restoring frictional force $F_t=k \Delta r_t$, up to a maximum force of $F_t=F_N$. The contact/friction model we use is relatively simple compared to some newer studies \cite{Jamali2019,More2020}. However, all these models exhibit qualitatively similar behavior, and the simpler model also compares favorably with experiments \cite{Lee2020}. In this study, we focus mainly on the differences between simulations with/without friction, rather than the quantitative details related to the exact frictional model chosen.

\section{Results}

All results are presented in reduced units, meaning distance is in units of particle diameter $d$, mass in units of particle mass $m$, and energy in units of $\varepsilon=k_BT$. The solvent is treated as a Newtonian fluid with viscosity $\eta_0=1$ in reduced units of $\sqrt{m\varepsilon}/d$, and all results on viscosity refer to the relative viscosity of the suspension compared to the solvent. Shear flow is applied in the $x$ direction, with gradient along $z$, and the results on 2D correlations or order parameters are in the flow-vorticity ($xy$) plane.

\subsection{Shear-induced ordering}
\label{sec:ordering}

\begin{figure*}
    \centering
    \includegraphics[width=.95\textwidth]{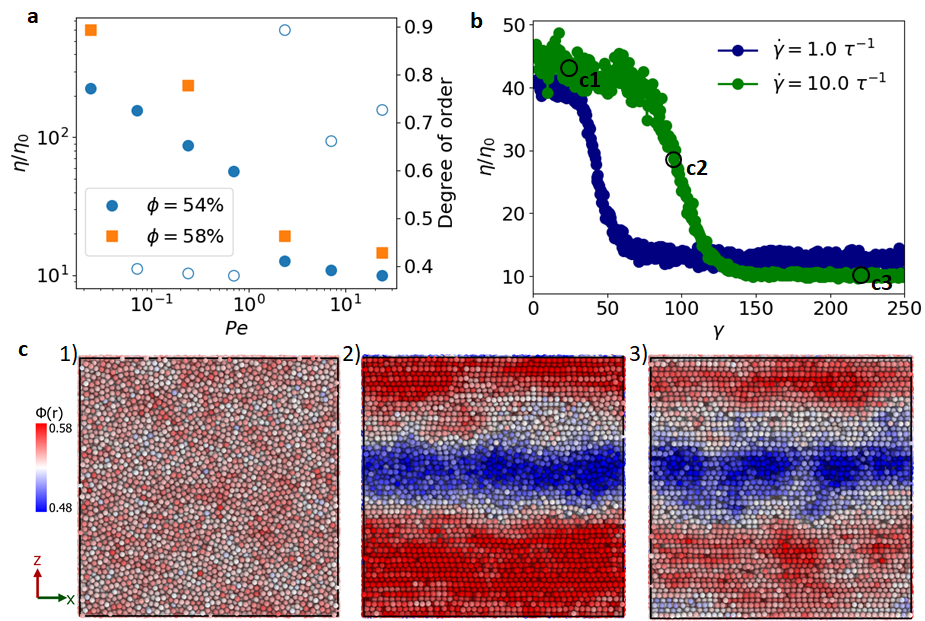}
    \caption{(a) The relative viscosity of the suspensions plotted as a function of the Peclet number displays shear-thinning behavior. There is an additional fairly sharp drop in the viscosity for $Pe>1$. The open symbols indicate the degree of order, which is defined as the fraction of particles with $\psi_6>0.6$ (discussed further in section~\ref{sec:orderingAnalysis}). For all points, data are averaged over a strain of $\gamma=5$ in steady state, and the standard deviation is smaller than the symbol size. (b) Plotting viscosity vs applied strains for two of these higher shear rates (with $Pe>1$) shows that there is a substantial drop in the viscosity at large $\gamma$. To help characterize this drop, we select three points (indicated by the labels c1, c2, and c3) along the shear curve for $\dot{\gamma}=10\,\tau^{-1}$ for further consideration. (c) Snapshots for the selected points, colored by the local volume fraction $\phi(r)$, computed over a small volume around each particle, show that the drop in viscosity is coupled to a change in the microstructure of the suspended particles.}
    \label{fig:viscosity}
\end{figure*}

We investigate the steady shear response of the model suspension of monodisperse colloidal spheres just described at volume fraction $\phi=54\,\%$. At this density, monodisperse hard sphere suspensions are in the liquid/crystal coexistance region of the phase diagram \cite{Pusey1986,Volkov2002}. However, our initial configuration is obtained by quenching the suspension to a metastable glassy state that does not exhibit any crystallization. We subject replicas of this configuration to steady shear at seven different shear rates (ranging from $10^{-2}\,\tau^{-1}$ to $10\,\tau^{-1}$). The Peclet number $Pe$, defined as the ratio of motion due to shear compared to motion due to Brownian forces, is $Pe=6\pi \eta_0 r^3 \dot{\gamma}/(k_BT)=\frac{3}{4} \pi \dot{\gamma}$ for the parameters chosen in our simulations.

As the simulation time is too short for crystallization to occur in the stationary suspension, the low $Pe$ flow exhibits a microstructure and viscosity that is independent of the elapsed time (and applied strain). The shear stress, and thus the apparent viscosity, fluctuate due to the Brownian motion of the suspended particles, but the steady-state is reached almost immediately. As shear rate is increased, we observe shear thinning behavior and the viscosity drops significantly as seen in Fig.~\ref{fig:viscosity}a. In addition, for $Pe>1$, there is a substantial difference in the initial viscosity and the steady-state viscosity.

For $\dot{\gamma}=1\,\tau^{-1}$ and $\dot{\gamma}=10\,\tau^{-1}$, the viscosity at the start of shear is about $\eta/\eta_0=45$, but as the applied strain increases the viscosity drops towards a final steady state value of $\eta/\eta_0 \simeq 10$ (Fig.~\ref{fig:viscosity}b). System snapshots are displayed in Fig.~\ref{fig:viscosity}c to show that this drop in $\eta/\eta_0$ is coupled to microstructural changes in the suspension. The pictured differences in local density are coupled to the formation of an ordered phase: the applied shear, at $Pe>1$, promotes a switch from the metastable glassy state to an ordered state. Signs of a similar transition have been seen in a previous study even with smaller applied strains \cite{Martys2012}.

\subsection{Analysis of ordering}
\label{sec:orderingAnalysis}

\begin{figure}
    \includegraphics[width=.49\textwidth]{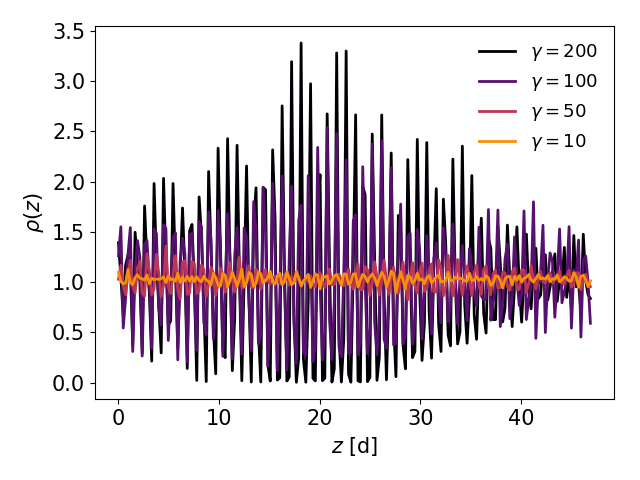}
    \caption{The density $\rho$ of particles in slices of the system along the gradient ($z$) direction at $\dot{\gamma}=10\,\tau^{-1}$ is plotted for various values of applied shear. The oscillations show that there is a shear-induced layering of the particles in the gradient direction. At low $\gamma$ the distribution is uniform and flat, but at high $\gamma$ there are many peaks corresponding to the positions of clearly distinguishable layers of particles. }
    \label{fig:densityProfile}
\end{figure}

We note that the ordering observed here is different from the equilibrium crystallization of similar suspensions at rest. As the system is sheared, the particles assemble into a layered configuration, with layers in the flow-vorticity ($xy$) plane spaced along the gradient direction ($z$). In Fig.~\ref{fig:densityProfile}, we show a plot of the local density $\rho(z)$. The density profile is flat at early times (low strains) but develops an oscillatory pattern at large $\gamma$. This shows the formation of particle layers, corresponding to the peaks in density, with center-to-center distance between the layers of $\simeq\,0.9\,d$. The shear flow imposes the requirement that these layers slide past each other, which disrupts the typical face-centered cubic (fcc) or hexagonal close-packed (hcp) crystal structure in the equilibrium crystals \cite{Pusey1986,Volkov2002}. Thus, the ordered state we observe consists of domains of stacked, 2D crystals coexisting with domains of lower density. To better characterize this microstructure, we focus our quantitative analysis on the structure within these layers and compute 2D correlation functions and order parameters.



\begin{figure*}
    \centering
    \includegraphics[width=.99\textwidth]{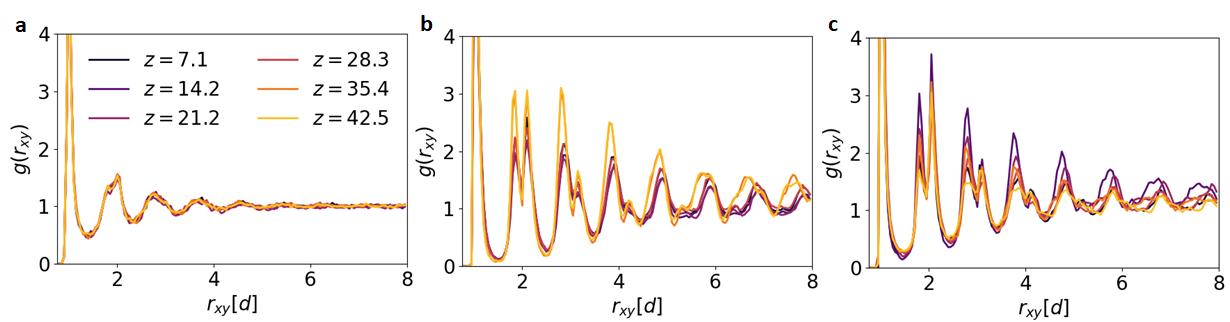}
    \caption{Pair correlation $g(r_{xy})$ in 2D layers computed in the $xy$ (flow-vorticity) plane. At low shear rates such as (a) $\dot{\gamma}=0.1\,\tau^{-1}$, $g(r_{xy})$ is independent of position within the system and shows a liquid-like order. Once steady state is reached for (b) $\dot{\gamma}=1\,\tau^{-1}$ and (c) $\dot{\gamma}=10\,\tau^{-1}$, there is a clear dependence of the structure on position in the suspension. Additionally, there are larger peaks in $g(r_{xy})$ indicating a more crystal-like ordering.}
    \label{fig:planarGr}
\end{figure*}

We first calculate a 2D version of the pair correlation function $g(r_{xy})$ in the flow-vorticity plane. This is computed within the previously described layers of particles, at various $z$ positions along the gradient direction, using the $xy$ distance between those particles. For the purposes of this calculation, particles $i$ and $j$ are considered to be in the same layer if $|z_i-z_j|<0.45\,d$, which is half the peak-peak distance in the oscillatory density profile seen in Fig.~\ref{fig:densityProfile}. The results of this calculation are shown in Fig.~\ref{fig:planarGr}, with plots of $g(r_{xy})$ for various values of $\gamma$ and $\dot{\gamma}$. 

In Fig.~\ref{fig:planarGr}a, we show the liquid-like ordering observed at lower rates (which is independent of $\gamma$ for the strains considered). There is a clear first neighbor peak, but the correlations rapidly decay beyond that. This also matches what we obtain from our initial state corresponding to a glassy, amorphous state. Note also that all the curves for different $z$ values lie on top of each other---the lack of ordering is consistent for all the layers along the gradient direction. 

Next, we consider the scenario at a higher rate ($\dot{\gamma}=1\,\tau^{-1}$). Here, the system transitions to a much lower viscosity state when sheared, accompanied by an increased layering of particles and the formation of domains with different local densities. In Fig.~\ref{fig:planarGr}b, we see that this substantially changes the 2D ordering of the particles. In particular, particle positions within layers are more strongly correlated and remain correlated over much larger distances. Moreover, the curves are no longer $z$-independent: the local structure varies across the system. At the highest shear rate of $\dot{\gamma}=10\,\tau^{-1}$, the peaks of $g(r_{xy})$ remain in similar positions and heights, but the variation with $z$ increases further (see Fig.~\ref{fig:planarGr}c).

\begin{figure}
    \centering
    \includegraphics[width=.45\textwidth]{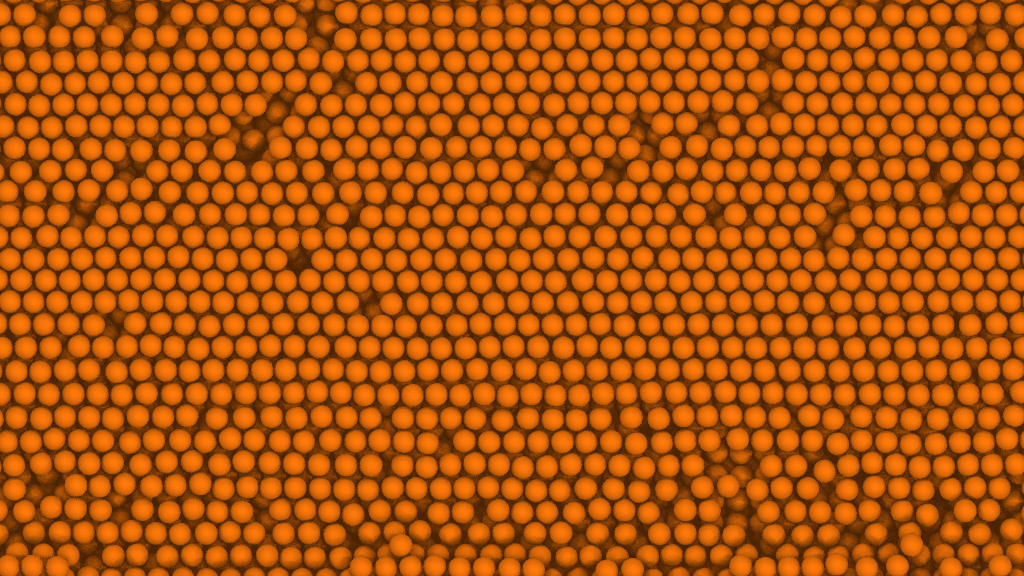}
    \includegraphics[width=.49\textwidth]{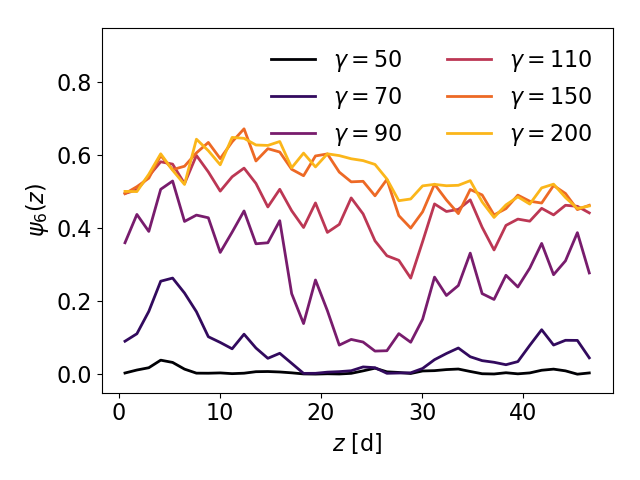}
    \caption{\textbf{Top:} Visualization of simulation snapshot of a layer of particles shows clear hexagonal packing in the flow-vorticity plane at $\dot{\gamma}=10\,\tau^{-1}$ and $\gamma=150$. \textbf{Bottom:} This is reflected in the average $\psi_6$ order parameter in layers along $z$, demonstrating the development of hexagonal crystal structure with shear. At low applied strain $\gamma$, the system remains disordered. As $\gamma$ increases, ordering starts to develop in parts of the system, and eventually a steady state with fairly high $\psi_6$ throughout the system is reached.}
    \label{fig:phi6}
\end{figure}

To better understand the changes in the local ordering, both in space and over time, we compute the 2D order parameter $\psi_6$ in the flow-vorticity plane. Fig.~\ref{fig:phi6} shows a snapshot of particles in the dense region of the sheared suspension at $\dot{\gamma}=10\,\tau^{-1}$ arranged in a hexagonal lattice, so the $\psi_6$ order parameter (which quantifies the degree of hexagonal order and goes to 1 in a perfect hexagonal lattice) is a natural choice to measure the local order \cite{Qi2006}. Like for the 2D pair correlations, we compute this parameter in the $xy$ plane for particles within a layer. $\psi_6^i$ is computed following Eq.~\ref{eq:phi6}. 

The particle order parameter $\psi_6^i$ can be averaged over the layer to obtain $\psi_6(z)$. This is the quantity we plot in Fig.~\ref{fig:phi6} for $\dot{\gamma}=10\,\tau^{-1}$ and different values of $\gamma$. By doing so, we see that the order parameter is uniformly $0$ in the disordered state at early times. As the strain accumulates, parts of the system begin to exhibit a sharp increase in $\psi_6$, indicating the nucleation and growth of ordered domains. The position of these ordered domains matches the higher local density regions seen in the snapshots of Fig.~\ref{fig:viscosity}, suggesting the ordering allows for more efficient packing of the particles. Also, as we saw for the density in those snapshots, the $\psi_6$ is highly heterogeneous when the structure first begins to change (and the viscosity begins to drop), but it becomes more uniform in the final steady state. The high amount of hexagonal order we observe in the steady state is in strong agreement with the scattering measurements of Lee et al \cite{Lee2018}, which reveal the same type of order in a shear-thinning colloidal silica suspension. A threshold of $\psi_6>0.6$ was used to determine the fraction of ordered particles in Fig.~\ref{fig:viscosity}a, and it is clear that this is much higher for $Pe>1$. Interestingly, as the shear rate is increased beyond that, there seem to be a small decrease in the fraction of particles exhibiting order, and in the next section we discuss the defects in the ordered states.

\subsection{Formation of defects}
\label{sec:defects}

\begin{figure}
    \centering
    \includegraphics[width=.45\textwidth]{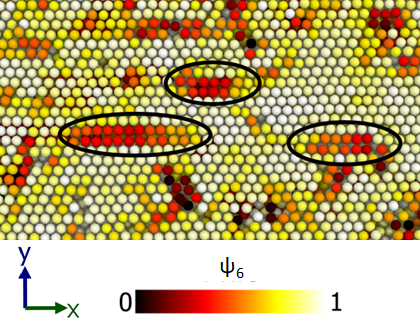}
    \caption{A snapshot of the flow-vorticity plane from simulations at $\dot{\gamma}=1\,\tau^{-1}$. The applied strain is $\gamma=150$, corresponding to the steady state, and particles are colored by their $\psi_6$ order parameter. We see predominantly hexagonal crystal order, and note that common defects in the hexagonal lattice (circled) correspond to chains-like structures of square lattice.}
    \label{fig:phi6vis}
\end{figure}

In steady state, with increased ordering and decreased viscosity, the relatively low average value of $\psi_6$ obscures the fact that the vast majority of particles have $\psi_6 \simeq 1$. In Fig.~\ref{fig:phi6vis}, we show a snapshot for $\dot{\gamma}=1\,\tau^{-1}$ where each particle is colored by its $\psi_6$ value. Here, it is clear that there are two populations: the majority of particles exhibit nearly perfect hexagonal order ($\psi_6 \simeq 1$), while there still exists a non-negligible minority that form domains of low $\psi_6$. This second population has a tendency to form chain-like structures where the hexagonal lattice is disrupted and the particles seem to arrange in a more square-like lattice. 

Evidence for chain structures, combined with hexagonal structures, has been seen before in the Stokesian dynamics simulations of Xu et al \cite{Xu2013}. However, given their limited system size ($<$ 200 particles) it was unfeasible to monitor the formation of these large-scale defects. Interestingly, similar defects have also been seen experimentally in particles assembled on a curved surface \cite{Irvine2010}. In that case, line defects, or ``pleats,'' formed as a way to handle the curvature-induced geometric frustration of the hexagonal packing. It is possible that we are observing a similar phenomenon but with stresses due to shear flow instead of curvature. If the square-like clusters are as common as this snapshot suggests, it should be possible to identify them with a different order parameter, $\psi_4$.

\begin{figure*}
    \centering
    \includegraphics[width=.9\textwidth]{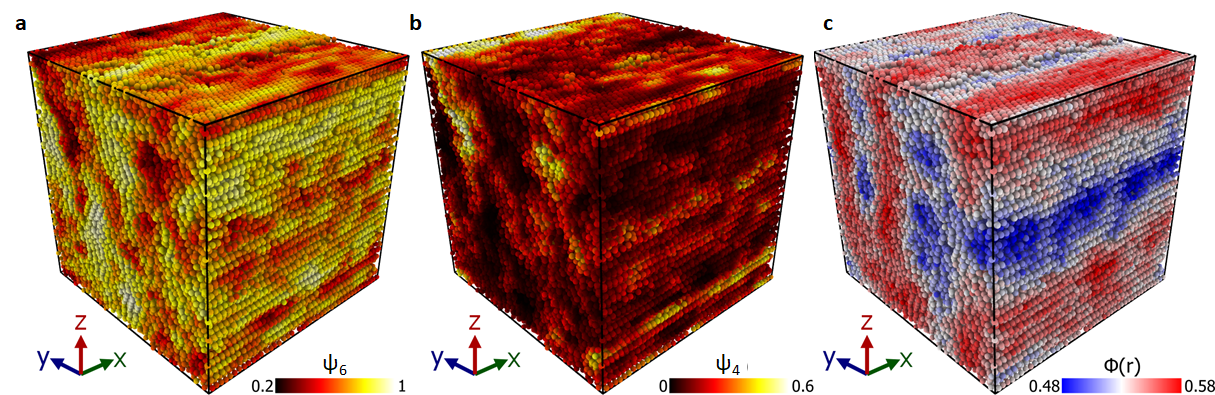}
    \caption{Visualizations of a steady-state configuration for $\dot{\gamma}=10\,\tau^{-1}$ with particles colored according to their (a) $\psi_6$ and (b) $\psi_4$ order parameter, coarse-grained over nearest neighbors. We observe that the two are anti-correlated with regions of high $\psi_6$ corresponding to low $\psi_4$, and vice versa. This is linked to the local density of the particles (c) since the hexagonal order allows for locally denser packing of particles.}
    \label{fig:phi6and4}
\end{figure*}

$\psi_4$, defined in Eq.~\ref{eq:phi4}, selects for 4-fold symmetry in the $xy$ plane, with nearest neighbors at 90 degree angles to each other. To suppress small fluctuations in the order parameter and highlight only larger groups of particles, we now average the particle order parameters over the nearest neighbor shells. This coarse-grained order parameter is shown in Fig.~\ref{fig:phi6and4}. Here the two snapshots correspond to the exact same configuration, colored by $\psi_6$ (left) and $\psi_4$ (right). By comparing these two, it becomes clear that the defects in the hexagonal crystal correspond to elongated domains of enhanced $\psi_4$. 

Crucially, these domains do not encompass entire layers of particles, which makes them difficult to detect via the quantitative layer analysis we performed earlier. However, the variations in the ordering are coupled to spatial fluctuations in the density. In the third snapshot of Fig.~\ref{fig:phi6and4}, we show the same configuration characterized by the local volume fraction $\phi=N_l V_p/V_l$, where $N_l$ is the number of particles within 5 particle diameters, $V_p$ is the volume of a particle, and $V_l=\frac{4}{3}\pi 5^3$. The larger regions of low $\psi_6$/high $\psi_4$ coincide with lower local density. Hexagonal order allows for denser packing of the spherical particles, but due to the shear flow a fully ordered state is not formed. Instead, we observe chain-like defects characterized by the $\psi_4$ order parameter and lower local density. As seen in Fig.~\ref{fig:viscosity}a, ordering is correlated with a substantial decrease in viscosity at $Pe>1$, but further increasing the shear rate decreases both the number of ordered particles and the viscosity. This implies that these elongated defects actually help to sustain steady, unimpeded flow. 

\subsection{Impact on stresses}
\label{sec:stresses}

\begin{figure*}
    \centering
    \includegraphics[width=.95\textwidth]{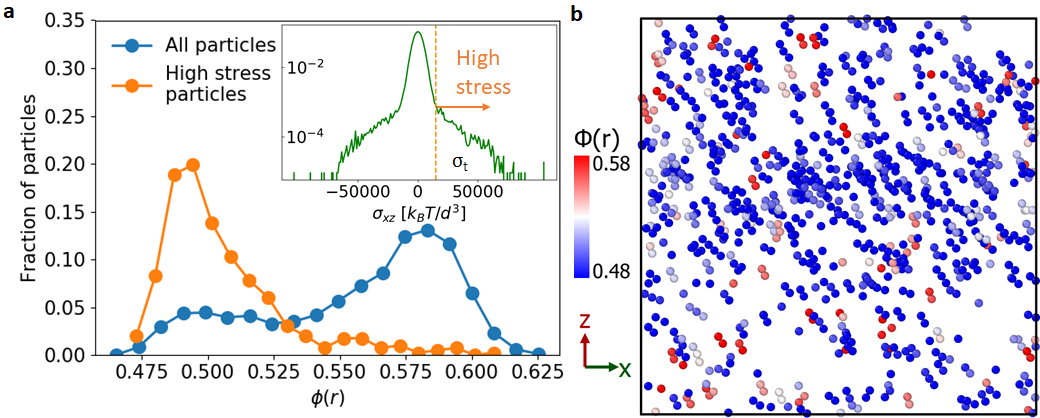}
    \caption{(a) The distribution of local volume fraction $\phi$ shows a bimodal distribution with a main peak around $\phi \simeq 0.58$, and a secondary peak at $\phi \simeq 0.49$. From the order parameter analysis, we know these two peaks correspond to the ordered regions and defects, respectively. A subset of particles that contributed the most to the shear stress is selected using Eq.~\ref{eq:stress} and setting a threshold of $\sigma^i_{xz}>\sigma_t=15,000 \, k_BT/d^3$, which corresponds to the top $1\,\%$ of particles, as shown in the inset. These high stress particles are predominantly situated in regions of lower density, and the local $\phi(r)$ distribution for these particles exhibits a peak at $\phi(r) \simeq 0.49$. (b) A snapshot, where only the high stress particles are rendered, shows that most of the stress is localized in low density regions where short chains of particles form in the compressive direction resisting shear. These stressed structures are dynamic, forming and breaking as particles rearrange in the defects.}
    \label{fig:stress}
\end{figure*}

Recognizing that all these microstructural changes (fluctuations in density, enhanced layering/ordering) are accompanied by a substantial drop in the viscosity of the suspension, we calculate the particle contributions to the stress tensor to quantify the link between the order/density and stresses \cite{Thompson2009}. The virial formulation of the stress tensor is broken up into particle contributions as:

\begin{equation}
    \sigma_{\alpha \beta}^i = -\sum_{j \neq i} \left[ \frac{1}{2} F_\alpha^{ij}r_\beta^{ij} \right] - mv_\alpha^i v_\beta^i
    \label{eq:stress}
\end{equation}
where $\alpha$ and $\beta$ can be $x$, $y$, or $z$ to generate the components of the stress tensor, $\vec{F}^{ij}$ and $\vec{r}^{ij}$ are the force and position vectors between particles $i$ and $j$, and $\vec{v}^i=\vec{v}^i_{\rm total}-\dot{\gamma}z \hat{x}$ is the deviation of the particle velocity from the flow profile set by the shear rate. The total stress tensor is obtained by summing $\sigma^i$ for all particles as in Eq.~\ref{eq:stressTotal}. The distribution of particle contributions to the shear stress is plotted in the inset Fig.~\ref{fig:stress}a, and it shows a peak near zero and long tails with a small fraction of particles contributing stresses that are very large in magnitude. This raises the question of which particles are responsible for these large stresses.

Computing the the local volume fractions in the neighborhood of each particle as before, we now analyze their distribution in Fig.~\ref{fig:stress}a. The blue curve shows a bimodal distribution, with the larger peak around $\phi\simeq 0.58$ and a smaller peak at $\phi\simeq0.49$. If we instead restrict our analysis to the high stress particles, selected by using a stress threshold of $\sigma_{xz}^i>\sigma_t=15000\,k_BT/d^3$ corresponding to $1\,\%$ of the particles, we see a single peak at the lower density. In other words, the high stress particles are almost exclusively in low density regions, i.e. the defects. This is represented visually in Fig.~\ref{fig:stress}b, where only the high stress particles are rendered. Notably, the high stress particles are almost all blue, corresponding to lower local volume fraction. In addition, they tend to form short chains in the compressive direction, which have been observed in previous simulations \cite{Lootens2008} and are highly reminiscent of the force chains in sheared granular materials \cite{Majmudar2005}. In our simulations, these chains form and break up rapidly as particles rearrange due to Brownian fluctuations and shear motion. Despite their transience, the presence of these structures shows a possible link between dense suspensions and granular systems even in the absence of friction.

At first, it seems somewhat counter-intuitive that higher stresses are associated with lower local density, but this reveals the tendency of the locally stressed structures to form in regions of defects. Yet, at a macroscopic level, we see that the presence of defects is driven by the applied shear rate (see fraction of ordered particles decreasing at $\dot{\gamma}>1\,\tau^{-1}$ in Fig.~\ref{fig:viscosity}a). Does stress cause defects or do defects cause stress? A possible explanation draws on previous work in 2D systems. The experiments of Irvine et al, for example, showed that hexagonally packed particles on curved surfaces formed chain-like defects as an optimal way to relax the curvature-induced stresses \cite{Irvine2010}. Additionally, the simulations of Schwenke et al showed that nanoparticles adsorbed to an interface formed a densely packed ordered phase with defects where adsorption could continue \cite{Schwenke2014}. Similarly, we observe growth of chain-like defects as we increase the macroscopic stress by increasing $\dot{\gamma}$. These defects may enable the larger ordered domains to slide past each other with minimal resistance. While they also allow for the formation of the compressed force chains seen in Fig.~\ref{fig:stress}b, those chains are transient as the contacts are lubricated and friction-less. In this way, the defects are an efficient response to the deformation-induced stresses, and their growth allows for a net decrease in the effective viscosity of the suspension. 

\subsection{Frictional forces disrupt order}
\label{sec:friction}

\begin{figure*}
    \centering
    \includegraphics[width=.95\textwidth]{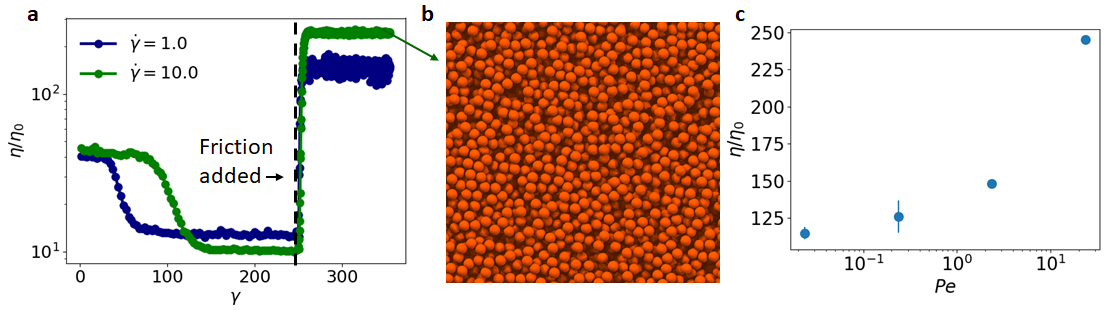}
    
    \caption{(a) As the suspension is sheared at $\dot{\gamma}=1\,\tau^{-1}$ and $\dot{\gamma}=10\,\tau^{-1}$, there is a large drop in shear stress due to ordering, and a steady state viscosity of $\eta \simeq 10$ is reached. Upon adding frictional forces, the stress rises rapidly to a new plateau corresponding to higher viscosity. (b) This new steady state exhibits none of the layering or hexagonal order found in the frictionless suspensions, as seen in a snapshot of the particles in the flow-vorticity ($xy$) plane. (c) Viscosity as a function of Peclet number, with data averaged over a steady-state strain of $\gamma=5$ (error bars indicate standard deviation). In addition to the disruption of ordering, the inclusion of frictional forces changes the rheological behavior and leads to shear thickening over the range of rates considered. }
    \label{fig:friction}
\end{figure*}

Thus far, all our results have been for entirely frictionless particles, but we have also investigated how the presence of friction might affect our findings. We use a frictional model based on the work of Mari et al \cite{Seto2013,Mari2014}, where a spring-like force acts against rotation of particles in contact. A key parameter of this model is the maximum frictional force that can be applied. In our simulations this has been set equal to the normal force between the particles, which places our simulations in the regime where shear thickening has been observed \cite{Seto2013,Mari2014}.

To test the effect of frictional forces on the ordering, we perform simulations with friction starting from the sheared ordered states discussed in previous sections. In Fig.~\ref{fig:friction}, we show results from simulations at four shear rates ($\dot{\gamma}=0.01\,\tau^{-1}$, $0.1\,\tau^{-1}$, $1\,\tau^{-1}$, and $10\,\tau^{-1}$) where the initial state has been sheared without frictional forces, meaning that for the high rate cases ($\dot{\gamma} \geq 1\,\tau^{-1}$) the system exhibits the layering and $\psi_6$ ordering discussed previously. 

In Fig.~\ref{fig:friction}a, we plot the shear stress as a function of applied shear strain. The first half shows the decrease in stress due to the ordering transition for frictionless spheres. After the dashed line, frictional forces are activated and there is a rapid increase in stress to a new plateau. This rise is correlated with the disruption of the ordered states. This disruption is very different from the defect regions discussed in previous sections which were still found to form layers. Fig.~\ref{fig:friction}b shows a snapshot from simulations at $\dot{\gamma}=10\,\tau^{-1}$, and the layered structure seen in the frictionless states is not preserved at all. 

In terms of the rheology, the addition of frictional forces dramatically increases the viscosity of the suspension and causes it to steadily shear thicken, with viscosity increasing by a factor of two over a decade increase in shear rate (see Fig.~\ref{fig:friction}c). With respect to the frictionless case, therefore, the rise in viscosity is associated with two factors: the extra stress from frictional forces and the disruption of the low-viscosity ordering. This behavior is in stark contrast with the thinning we observed without friction (see Fig.~\ref{fig:viscosity}), and in agreement with the results of previous studies which indicate that friction is the key to shear thickening \cite{Seto2013,Mari2014,Wyart2014,Singh2020,Morris2018,Royer2016,Pradeep2021}. 

\begin{figure*}
    \centering
    \includegraphics[width=.95\textwidth]{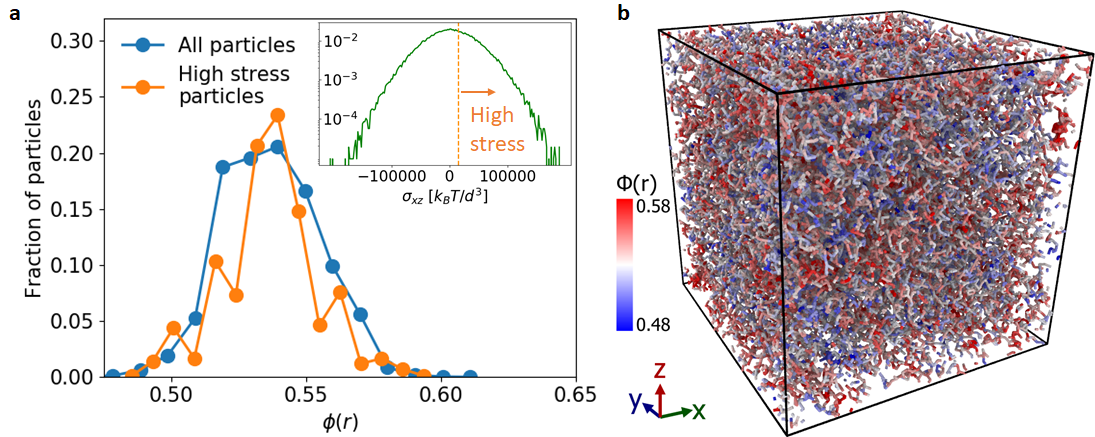}
    
    \caption{(a) The particle histogram of local volume fraction $\phi(r)$ at $\dot{\gamma}=10\,\tau^{-1}$ shows a single peak in the frictional simulations, highlighting the lack of coexisting ordered and defected regions. In addition, the particle stresses (inset) are distributed normally, without long tails. We perform the same analysis and consider particles selected so that $\sigma_{xz}^i>\sigma_t=15,000 \, k_BT/d^3$ (the same criteria used for the frictionless case). With friction, $\simeq 35\,\%$ of particles meet the stress critieria, and those particles are distributed uniformly throughout the system in both high and low $\phi$ regions. (b) A snapshot of the high stress particles, colored by the local volume fraction, illustrates that with frictional forces the stress is distributed throughout the system rather than only in a few small regions of defects. For this visualization, only the contacts between particles are rendered in order to highlight the formation of a connected and system-spanning network. The pattern of red/blue coloring shows that there are still interesting structural features, but the high stress network is somehow embedded within, and spanning across, those features.}
    \label{fig:friction2}
\end{figure*}

Analysis of the particle contributions to stresses in our system shows that there is a fundamental change in which particles contribute to the increased shear stress. Mirroring the analysis in section~\ref{sec:stresses}, we show the particle histogram of local $\phi$ in Fig.~\ref{fig:friction2}a. The presence of a single peak indicates that we no longer have two coexisting states (order/defect) as in the frictionless simulations. Additionally, the particle contributions to the shear stress are distributed normally, and using the same criteria for high stress particles, $\sigma_{xz}^i>\sigma_t=15000\,k_BT/d^3$, now selects $\simeq 35\,\%$ of the particles instead of just $1\,\%$. Interestingly, the histogram for the local volume fraction of these particles matches that for all particles, indicating that the link between local volume fraction and local stress is much less obvious here.

Visualizing the contacts between particles as bonds (Fig.~\ref{fig:friction2}b), we find that the high stress particles form a percolating network that spans both lower (blue) and higher (red) $\phi$ regions. The ability of stresses to accumulate in a large-scale structure, instead of only in highly localized regions of defects, explains the substantial increase in the suspension viscosity. The existence of a percolated stress/contact network matches previous simulation studies of frictional systems \cite{Seto2013,Mari2014,Morris2018}. 

We find that the inclusion of friction has not only changed the microstructure of the suspension, it has also changed how stress is carried within that structure. Instead of the majority of the stress response arising from small regions of defects, stress can accumulate in a space-spanning structure. The formation of this larger-scale, percolating network structure raises some interesting new lines of inquiry. For one, studying the heterogeneity of the network, in terms of structure and local stresses, could provide new insight into the shear thickening behavior and the stress fluctuations seen in experiments \cite{Rathee2020,Ovarlez2020}. Similarly, the dynamic evolution of the network under shear is another avenue to explore in future work. A better understanding of the dynamics would assist in efforts to tune the thickening, perhaps through perturbations like those by Sehgal et al \cite{Sehgal2019}.

\section{Conclusions}
We performed simulations of a dense colloidal suspension undergoing shear. The simulation approach included hydrodynamic lubrication forces and short-ranged repulsive interactions. The use of LAMMPS allowed us to perform large-scale simulations with $10^5$ particles in each sample. At high shear rates, $Pe>1$, we find that the shear flow induces an ordering transition in the suspension that led to a significant reduction in the overall viscosity. Shear-induced crystallization has also been seen previously in experiments \cite{Lee2018,Ackerson1990,Mabhoff2020} and simulations \cite{Rastogi1996,Kulkarni2009,Stevens1993} of colloidal suspensions, as well as in polymer solutions \cite{Nguyen2020}. By computing 2D pair correlations and the $\psi_6$ order parameter, we have determined that the ordered state consisted of hexagonally packed layers that slide past each other as the system is sheared in steady state. Due to our use of a large system size, we were able to observe the nucleation and growth of large ordered domains which coexist with lower density defect regions. The defects typically formed chain-like structurues along the flow direction and can be identified by the $\psi_4$ order parameter. Surprisingly, these lower density regions actually contribute the most to the overall stress response due to the formation of short, transient chains of particles resisting compression. These ``force chains'' are dynamic, rapidly forming and breaking apart under shear, but are reminiscent of the force chains observed in jammed granular systems. The fact that these are mainly restricted to the defected regions could explain why the overall shear stress decreases when the ordered phase forms. 

Upon introducing frictional forces between particles after reaching the ordered state, we observe that the ordering is rapidly disrupted. Friction changes the rheological properties of the suspension by drastically increasing viscosity and causing the suspension to shear thicken instead of shear thin. Our analysis of the particle contributions to the stress shows that these phenomena are accompanied by a major redistribution of the stresses within the microstructure. Instead of being localized in less dense regions as in the frictionless case, the particles with large stress contributions form a percolating network that spans low and high density regions. It is possible that the additional constraints on particle motion due to friction may serve to stabilize and grow the stress-carrying structures that were formed in the frictionless case \cite{Vinutha2016}, thus disrupting the order and leading to the observed network. These results show how the nature of particle contact, for otherwise identical suspensions, can control the transition from steady state order to disorder, or thinning to thickening. Additionally, they shed new light on how the link between particle interactions and microstructure controls both the stress distribution and the macroscopic flow properties. 

\section{Acknowledgements}
The authors acknowledge the NIST PREP Gaithersburg Program (70NANB18H151), National Science Foundation (NSF DMR-2026842), Georgetown University for support.

\bibliographystyle{ieeetr}
\bibliography{library}

\end{document}


\title{Supporting Information}

\maketitle

\section{3D order parameters}

To characterize the 3D ordering, we looked at the Steinhardt Bond Orientational Order Parameters (BOOPs) \cite{Steinhardt1983}. These are computed by looking at angular orientation of "bonds" (not chemical bonds, but rather bonds defined by a suitable nearest neighbor selection). For a bond with some orientation $\vec{r}$, we consider the spherical harmonic function, $Y_{lm}(\theta(\vec{r})\phi(\vec{r}))$. The BOOPs, $q_{lm}$, are defined as an average over the bonds of a particle:
	\begin{equation}
	q_{lm}(i) = \frac{1}{N_{b_i}} \sum_{j=1}^{N_{b_i}} Y_{lm}(\theta(\vec{r_{ij}})\phi(\vec{r_{ij}}))
	\end{equation}
	We are interested in the even $l$ terms, which are independent of an arbitrary choice of bond direction. These terms are still dependent on the choice of reference frame though. To remove this dependence, we must consider rotationally invariant combinations of $q_{lm}$. The second-order rotational invariants are
	\begin{equation}
	q_l(i) = \sqrt{  \frac{4\pi}{2l+1}  \sum_{m=-l}^{l}  |q_{lm}(i)|^2  } 
	\end{equation}
	and the third-order invariants are
	\begin{equation}
	\hat{w}_l(i) = w_l(i)/ \left[  \sum_{m=-l}^{l}    |q_{lm}(i)|^2    \right] ^{3/2}
	\end{equation}
	where
	\begin{equation}
	w_l(i)= \sum_{m_1,m_2,m_3} \binom{l\ \ \ \ l\ \ \ \ l}{m_1\ m_2\ m_3} q_{lm_1}(i) q_{lm_2}(i) q_{lm_3}(i)
	\end{equation}
	where the coefficients are the Wigner 3-j coefficients and the sum is over values such that $m_1+m_2+m_3=0$. Specifically, the set of $\left\lbrace  q_4, q_6, \hat{w}_4, \hat{w}_6 \right\rbrace $ are generally sufficient to characterize the order of colloidal systems.
	
	\begin{figure}
	    \centering
	    \includegraphics[width=.95\textwidth]{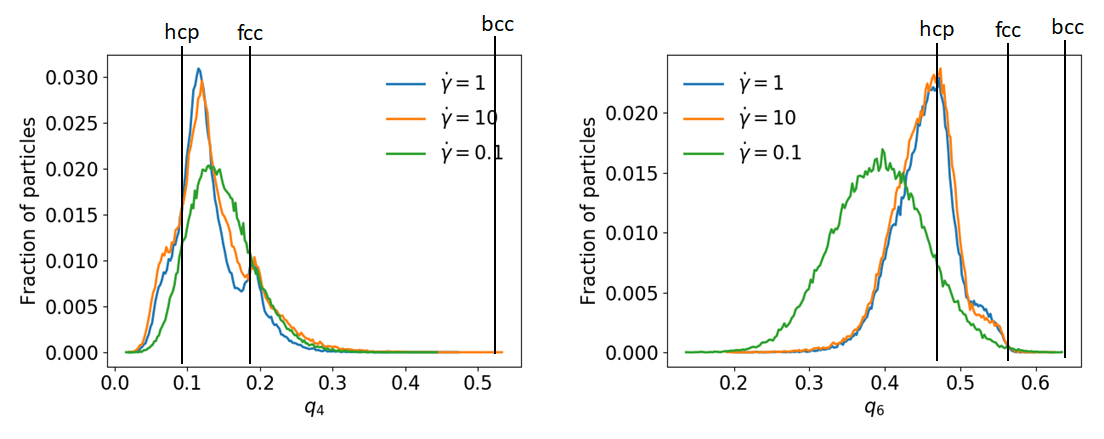}
	    \caption{The particle distributions of $q_4$ and $q_6$ BOOPs. The vertical lines indicate the characteristic values associated with various types of crystals: face-centered cubic (fcc), body-centered cubic (bcc), and hexagonal close-packed (hcp). }
	    \label{fig:boopq}
	\end{figure}
	
	\begin{figure}
	    \centering
	    \includegraphics[width=.95\textwidth]{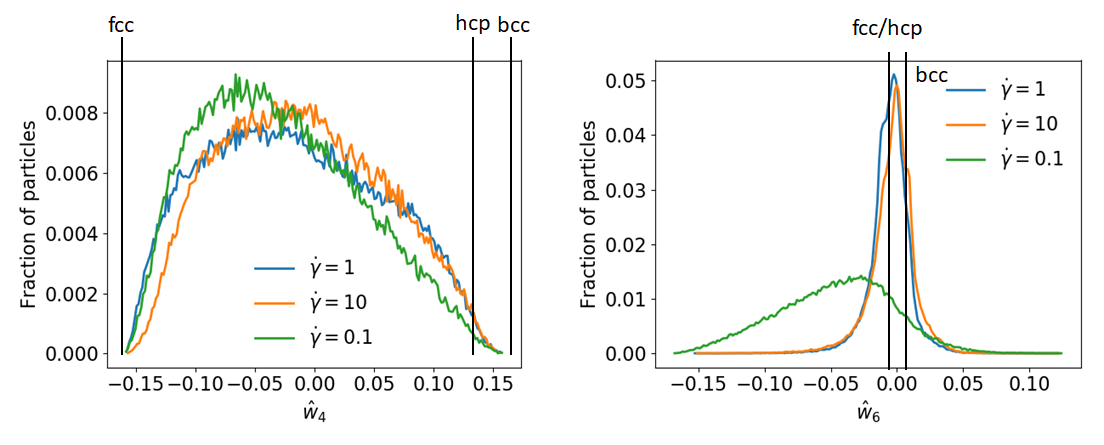}
	    \caption{The particle distributions of $\hat{w}_4$ and $\hat{w}_6$ BOOPs. The vertical lines indicate the characteristic values associated with various types of crystals: face-centered cubic (fcc), body-centered cubic (bcc), and hexagonal close-packed (hcp).}
	    \label{fig:boopw}
	\end{figure}
	
	In Fig.~\ref{fig:boopq} and Fig.~\ref{fig:boopw}, we show the distributions of each computed BOOP. The vertical lines indicate characteristic values for 3D crystal structures, and taken together it is evident that none of those crystals is a match for the ordering observed. There is certainly a clear difference between the BOOP distributions for $\dot{\gamma}=0.1$ compared to $\dot{\gamma}=1$ and $\dot{\gamma}=10$, which reflects the ordering transition discussed in the main text. However, the fact that the suspension is being sheared makes it impossible to form a full 3D crystal structure, so we focus our discussion in the main paper on 2D characterization of the order.
	
	\bibliography{library}